\documentstyle[12pt,epsf]{article}
\begin{document}
\begin{titlepage}
\begin{center}
{\Large\bf Instanton tunneling and 
multiphonon giant resonances \\}
\vspace*{15mm}
{\bf T. Obikhod\\}
\vspace*{3mm}
{\it Institute for Nuclear Research, National Academy of
Sciences of Ukraine\\
252022 Kiev, Ukraine\\}
e-mail: malyuta@olinet.isf.kiev.ua\\
\vspace*{7mm}
{\bf S. Radionov\\}
\vspace*{3mm}
{\it Physics Department of Kiev State University\\
252028 Kiev, Ukraine\\}
\vspace*{30mm}
{\bf Abstract}
\end{center}
\large
\vspace*{1mm}
\hspace*{7mm}The sine-Gordon model is applied to 
describe the multiphonon
giant resonances. 

\end{titlepage}
\newpage
\large
\hspace*{-7mm}Direct observations of the multiphonon giant 
resonances in $ ^{208} $Pb and $ ^{136} $Xe were obtained
at the GSI/SIS, Darmstadt \cite{1.,2.}.
Centroid energies and widths of these resonances are
given in Table 1.
\begin{center}
\vspace*{7mm}
Table 1\\
\vspace*{5mm}
\begin{tabular}{|ccc|}          \hline  \hline
              &         &             \\
Nucleus       &\hspace*{6mm} Centroid energies  &  Widths  \\
              &         &     \\      \hline  
              &         &     \\ 
$ ^{208} $Pb &\hspace*{5mm} $ {{E_{2}}\over{E_{1}}} $ = 
$ 1.93\pm0.07  $  &
\hspace*{5mm} $ {{\Gamma_{2}}\over{\Gamma_{1}}} $ = 
$ 1.4 \pm 0.4 $ \\ 
              &         &     \\      \hline
              &         &     \\
$ ^{136} $Xe &\hspace*{5mm} $ {{E_{2}}\over{E_{1}}} $ = 
$ 1.86\pm0.05 $  &
\hspace*{5mm} $ {{\Gamma_{2}}\over{\Gamma_{1}}} $ = 
$ 1.3 \pm 0.4 $  \\ 
       &                &     \\      \hline \hline
\end{tabular}
\end{center}  
\vspace*{1cm}
\hspace*{7mm}The purpose of this paper is to describe 
the multiphonon giant resonances 
in terms of the sine-Gordon model.

	The Lagrangian of the sine-Gordon model 
has the form \cite{3.}
\[  L=-{{1}\over{2}}({{dq}\over{d\tau}})^2-V(q) 
\hspace*{3mm} ,  \]
where 
\[ V(q)={{a}\over{b}}(1-\cos bq) \] 
is the potential, $ q $ is a 
generalized coordinate, $ \tau $ is an imaginary 
time, $ a $ and $ b $
are parameters. The potential $ V(q) $ 
is shown in Figure 1, where
$ q_{n}=2\pi n \omega/ \sqrt{\lambda} $ is the 
coordinate of the n-th
well, $ \omega=\sqrt{ab} $ is the phonon frequency, 
$ \lambda=ab^{3} $ is the constant of anharmonicity.\\
\begin{center}
\begin{tabular}{c}
\vspace*{2mm}\\
\epsfxsize=6cm\epsffile{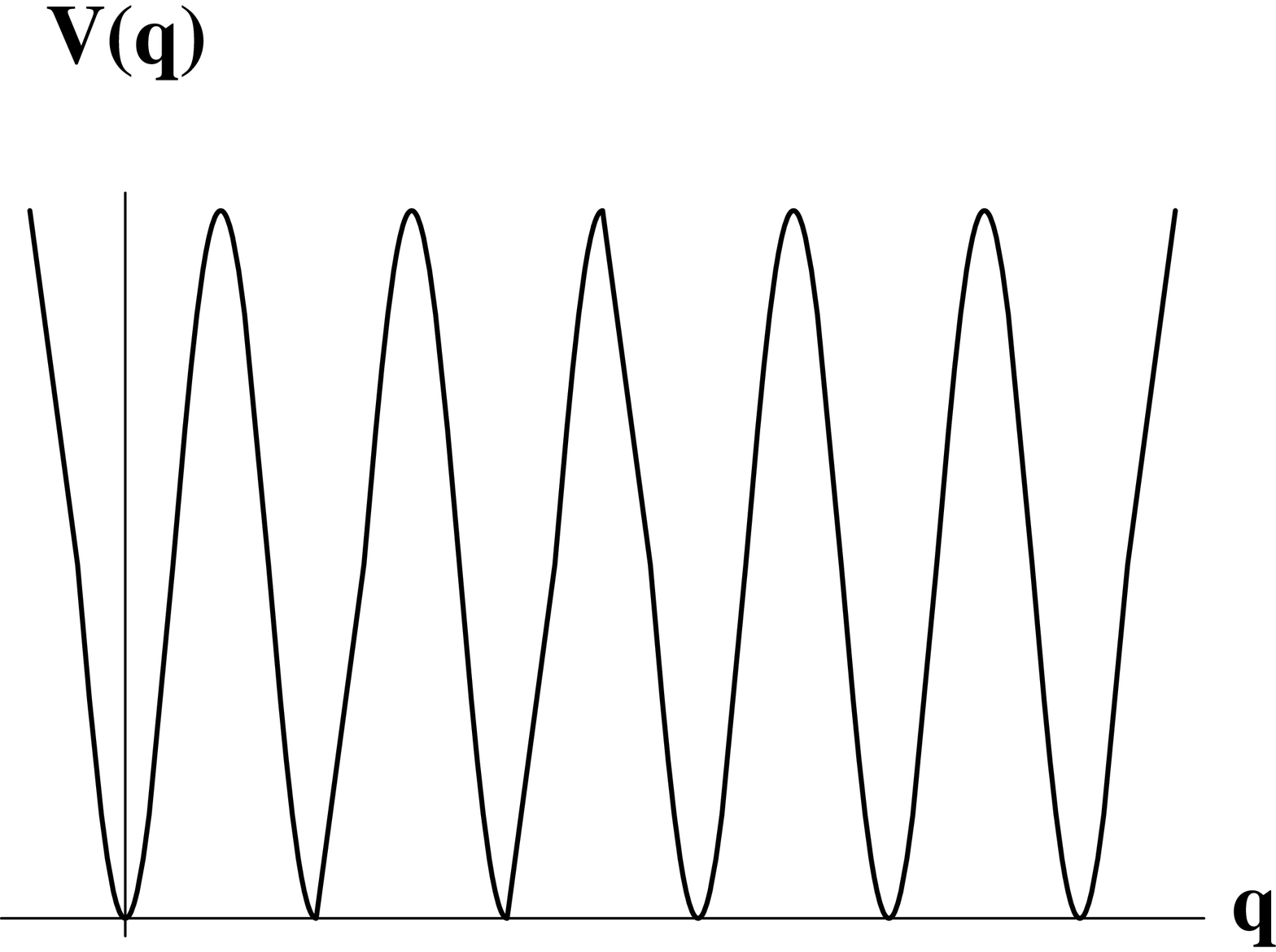}
\vspace*{5mm}\\
Figure 1\\
\end{tabular}
\end{center}
\vspace*{7mm}
Parameters $ \omega $ and $ \lambda $ appear in the 
Taylor expansion
\[ V(q)={{\omega^{2}}\over{2}} q^{2}-{\lambda\over{4!}}q^{4}+
\ldots \hspace*{3mm}, \]
where first two terms represent the anharmonic 
oscillator potential.
\hspace*{7mm}The transition amplitude between 
multiphonon states 
$ |q_{n-1}> $ and $ |q_{n}> $ is defined by the 
Feynman path integral \cite{4.}
\begin{equation}
<q_{n}|e^{-H\tau}|q_{n-1}>=\int[dq]e^{-S[q]} \hspace*{3mm}, 
\end{equation}
where $ S[q]=\int Ld \tau $ is the action, $ e^{-H\tau} $ 
is the
evolution operator, $ H $ is the Hamiltonian.\\
\hspace*{7mm}Using the stationary-phase approximation 
\cite{4.}
for calculation of the path integral (1) we find 
\begin{equation}
<q_{n}|e^{-H\tau}|q_{n-1}>=\sqrt{{\omega}\over{\pi}}
e^{-S[q_{cl}]} 
\hspace*{3mm},
\end{equation}
where the classical coordinate $ q_{cl} $ satisfies 
the following
condition \cite{3.}
\[ H=-{{1}\over{2}}({{dq}\over{d\tau}})^2+V(q)=0 
\hspace*{3mm}. \]
The classical action is given by
\[ S[q_{cl}]=
\int\limits_{2\pi n/b}^{2\pi(n-1)/b}{\sqrt{2V(q)}}dq 
={8{\omega}\over{b^{2}}} \hspace*{3mm}. \]
The resonance width, which correspond to the transition (2)
is equal to
\begin{equation}
\Gamma_{n}=|<q_{n}|e^{-H\tau}|q_{n-1}>|^{2}=
{{\omega}\over{\pi}}
e^{-16\omega/b^{2}} \hspace*{3mm}.
\end{equation}
From (3) follows the relation 
\[ {{\Gamma_{n}}\over{\Gamma_{n-1}}} = 1 \hspace*{3mm}.  \]
The experimental data from Table 1 satisfy this relation
within experimental errors.\\
\hspace*{7mm}Note also that the sine-Gordon model 
describes centroid energies by the relation
\[ {{E_{n}}\over{E_{n-1}}} = {{n}\over{n-1}} 
\hspace*{3mm}. \]
This relation agrees with the experimental data from 
Table 1.\\ 
\vspace*{6mm}\\
\hspace*{7mm}ACKNOWLEDGEMENT: We are pleased to thank 
Hans Emling for
sending us the experimental data.

\end{document}